\documentclass[]{spie}  
\usepackage{graphicx} 

\usepackage{subfigure}
\usepackage{float}
\usepackage{siunitx}
\usepackage{multicol}
\usepackage[table]{xcolor}

\graphicspath{ {immagini/} } 

\title{A virtual coronagraphic test bench for SHARK-NIR, the second-generation high contrast imager for the Large Binocular Telescope}

\author{
Vassallo D.\supit{a,b,f}, 
Carolo E.\supit{a,f}, 
Farinato J.\supit{a,f}, 
Agapito G.\supit{d,f},
Bergomi M.\supit{a,f},
Carlotti A.\supit{c},
De Pascale M.\supit{a},
D'Orazi V.\supit{a,f},
Greggio D.\supit{a,f},
Magrin D.\supit{a,f},
Marafatto L.\supit{a,f},
Mesa D.\supit{a,f},
Pinna E.\supit{d,f},
Puglisi A.\supit{d,f},
Stangalini M.\supit{e,f},
Verinaud C.\supit{c},
Viotto V.\supit{a,f},
Biondi F.\supit{a,f},
Chinellato S.\supit{a,f},
Dima M.\supit{a,f},
Esposito S.\supit{d,f},
Pedichini F.\supit{e,f},
Portaluri E.\supit{a,f},
and Ragazzoni R.\supit{a,f}
\skiplinehalf
\supit{a}INAF - Osservatorio Astronomico di Padova, vicolo dell'Osservatorio 5, 35141 Padova, Italy; \\ 
\supit{b}Dipartimento di Fisica e Astronomia, Universit\`a degli Studi di Padova, Vicolo dell'Osservatorio 3, 35122, Padova, Italy; \\
\supit{c}Institut de Plan\'etologie et d'Astrophysique de Grenoble, 414 Rue de la Piscine, Domaine Universitaire, 38400 St-Martin-d'Hères, Grenoble, France; \\
\supit{d}INAF - Osservatorio Astrofisico di Arcetri, Largo Enrico Fermi 5, 50125 Firenze, Italy; \\
\supit{e}INAF - Osservatorio Astronomico di Roma, Via Frascati 33, 00078 Monte Porzio Catone, Roma, Italy; \\
\supit{f}ADONI - Laboratorio Nazionale Ottiche Adattive - National Laboratory for Adaptive Optics, Italy 
}

\authorinfo{Further author information: (Send correspondence to D.V.)\\D.V.: E-mail: daniele.vassallo@oapd.inaf.it, Telephone: +393285771394} 

\begin{document} 
  \maketitle

\begin{abstract}
In this article, we present a simulator conceived for the conceptual study of an AO-fed high-contrast coronagraphic imager. The simulator implements physical optics: a complex disturbance (the electric field) is Fresnel-propagated through any user-defined optical train, in an end-to-end fashion.\\The effect of atmospheric residual aberrations and their evolution with time can be reproduced by introducing in input a temporal sequence of phase screens: synthetic images are then generated by co-adding instantaneous PSFs. This allows studying with high accuracy the impact of AO correction on image quality for different integration times and observing conditions.\\In addition, by conveniently detailing the optical model, the user can easily implement any coronagraphic set-up and introduce optical aberrations at any position. Furthermore, generating multiple images can allow exploring detection limits after a differential post-processing algorithm is applied (e.g. Angular Differential Imaging).\\The simulator has been developed in the framework of the design of SHARK-NIR, the second-generation high contrast imager selected for the Large Binocular Telescope.
\end{abstract}



\section{INTRODUCTION}
\label{sec:intro}  
SHARK-NIR is a coronagraphic camera. Since coronagraphs deal with diffraction of light, it is necessary to operate in the framework of wave-propagation physics: given the mathematical complexity of this theory, the most common approach is to make use of numerical simulations. The test bench is written in IDL language and it is developed to assess the coronagraphic performance of the camera.\\Section \ref{sec:SC} describes the simulator, with a brief insight into critical aspects such as implementation of AO correction and NCPA. Section \ref{sec:coro} introduces the coronagraphic techniques implemented in the test bench and identified as possible candidates for SHARK-NIR. Finally, section \ref{sec:results} shows some of the possible studies that can be performed with the simulator.

\section{The Simulator}
\label{sec:SC}
The tool generates synthetic coronagraphic images using Fresnel optical propagation. For this purpose, we chose the IDL library PROPER\cite{PROPER}. The routines of this library allow to propagate numerically an electric field through an optical train according to scalar theory of diffraction, in an end-to-end fashion. The propagator (i.e. numerical alghoritm) to move from one surface to the subsequent one is selected by a dedicated set of internal routines by means of analytical propagation of a Gaussian pilot beam, resulting in a very accurate modeling of the diffraction phenomenon. Finally, intrinsic limitations of physical optics propagation with respect to ray tracing do not play an important role because of the absence of refractive or highly aspherical optics in the camera, together with the excellent optical quality by design.\\

\subsection{AO correction and computational time}
LBT AO system FLAO (First-Light AO)\cite{FLAO} is a Natural Guide Star (NGS) single-conjugate system whose key strengths are the telescope adaptive secondary mirror\cite{M2} and the pyramid wavefront sensor\cite{PIRAMIDE}. Since the efficiency of the AO system feeding the scientific instrument is one of the fundamental drivers of coronagraphic performance, a realistic estimate of the correction delivered by FLAO is mandatory. For the purpose, as input the code uses closed-loop phase residuals generated with the official FLAO simulator PASSATA\cite{PASSATA}. This code has been developed at INAF-Arcetri (Florence) and has been proven to yield SR in very good agreement with on-sky measurements.\\
The framerate of the AO loop sets the time step of simulations. For targets brighter then $R\sim10$, FLAO delivers wavefront correction at 1 kHz rate, meaning that phase screen are interleaved by 1 millisecond.\\
Following this approach, the simulation time required to generate an image of exposure $t_{exp}$ can be expressed as: 
$$T=T_{0}  \frac{t_{exp}}{\Delta t}$$ 
namely, a multiple of the time $T_{0}$ required for the numerical propagation through the whole optical train, which is primarily dependent on the size of computational matrices and on the amount of intermediate surfaces in the model of the camera. $\Delta t$ is the time step discussed before.\\
The code also implements an optimized parallelization scheme, which allows to reduce this time by a factor depending on the number of simultaneous threads the machine is able to handle. The final result in our simulations is $\sim10$ minutes (computation time) per second of integration at 1 kHz rate.\\

\subsection{Non Common Path Aberrations}
\label{sec:NCPA}
Speckles originating because of Non Common Path Aberrations (NCPA) represent a significant source of noise in high contrast imaging because of their long life and slow temporal evolution. These aberrations are introduced in the virtual test bench, according to some simplifying assumptions: we assumed a flat power spectral density function (PSD) until the AO cut-off frequency, followed by a $f^{-2}$ decline. This model relies on the assumption that the introduced NCPA are actually residuals aberrations after a compensation is performed with LBT Adaptive Secondary Mirror (ASM). In addition, we assumed a total power of 30 nanometers rms. This value is the result of a detailed assessment of NCPA error budget for SHARK-NIR and of a dedicated study on the possible compensation strategies.\\ 

\subsection{Telescope vibrations}
\label{sec:jitter}
Vibrations arise from resonant modes in the structure of the telescope, in particular the swing arm supporting the ASM. These modes are excited by wind shacking and/or telescope tracking and mainly introduce tip and tilt aberrations which are only partially filtered out by the AO system. Vibrations at LBT have been characterized during FLAO commissioning run: the median value is around $6$mas rms, with most of the power concentrated at a frequency of $\sim13$ Hz. In order to model vibrations realistically, we used on-sky data coming from the Forerunner experiment, a high contrast experimental imager installed at LBT in 2014\cite{FORERUNNER}. The data set consists of a sequence of more than one million frames acquired at 1 ms cadence. The vector of centroids measured on these images can be easily transformed into a tip-tilt aberration to reproduce in simulations the same star `wandering'. Additionally, data can be arbitrarily rescaled in order to reproduce any total vibration power but preserving its temporal properties. 

\section{Coronagraphs}
\label{sec:coro}
Thanks to the presence of two intermediate pupil planes and one focal plane in between them, the optical design of SHARK-NIR is compatible with several coronagraphic techniques. From Lyot-class coronagraphs, requiring only the focal plane and the downstream pupil plane, to three-planes techniques exploiting pupil apodization, like Shaped Pupil and Apodized Pupil Lyot Coronagraph. In this section we introduce all coronagraphs implemented in the test bench.  

\subsubsection{Gaussian Lyot coronagraph}
\label{sec:Lyot}
This technique belongs to the class of Lyot coronagraphs, consisting of an amplitude mask in the focal plane and a stop in the downstream pupil plane (the Lyot stop) to mask the light diffracted at the edge of the pupil itself. In this variant, the amplitude of the electric field in the focal plane is modulated with a Gaussian filter. SHARK-NIR will implement this solution for its overall robustness. Simulations showed that it represents an optimal choice to adress those science cases not requiring extreme contrast nor very small IWA. The current design features a $3$ $\lambda$/D occulter and a downstream pupil undersizing of 15\%.

\subsubsection{Shaped Pupil}
\label{sec:SP}
\begin{figure}[h]
\label{fig:SP_example}
\centering
{\includegraphics[width=.40\columnwidth]{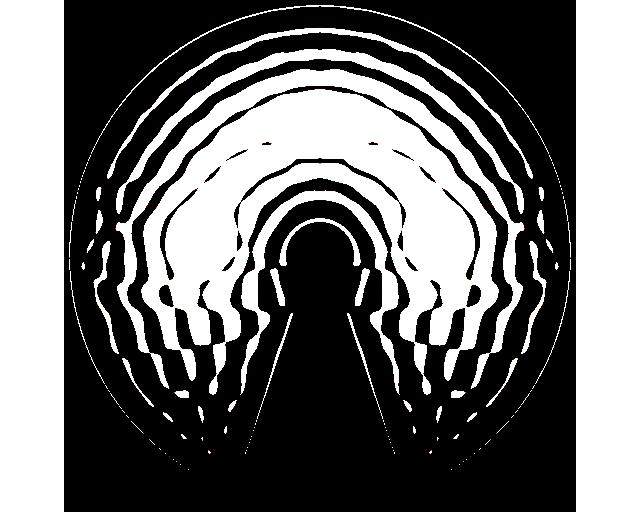}} \quad
{\label{fig:example-b}%
\includegraphics[width=.40\columnwidth]{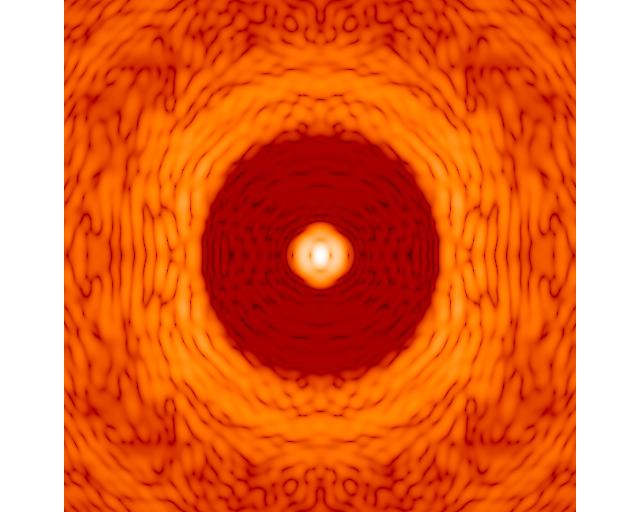}} \\ \vspace{0.2 cm}
{\includegraphics[width=.45\columnwidth]{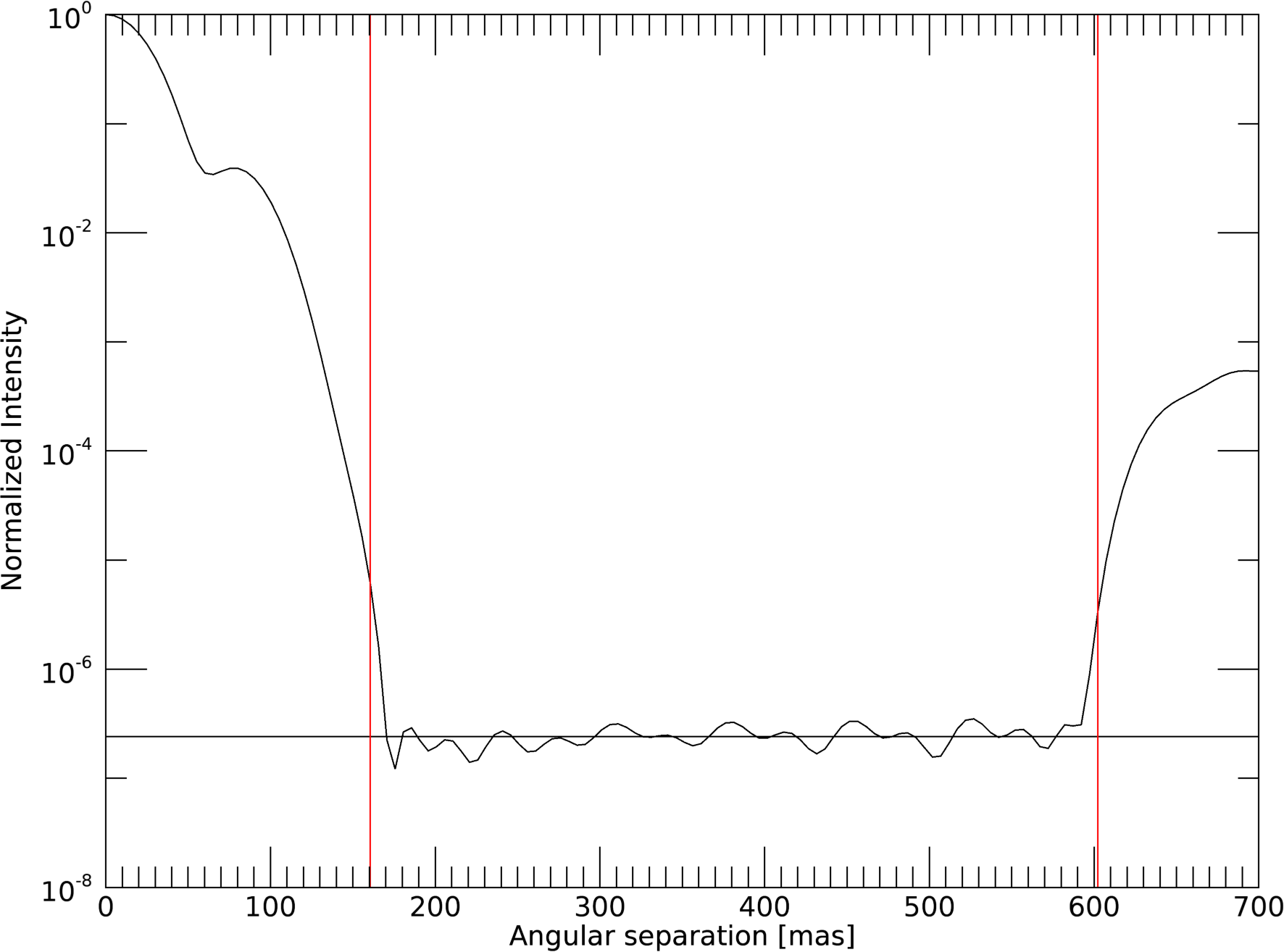}} \quad \vspace{0.3cm}
\caption{Example of Shaped Pupil design. \textit{Top-left}: the binary pupil mask. \textit{Top-right}: Intensity distribution in the focal plane (Fourier transform of the re-shaped pupil). \textit{Bottom:} Radial intensity profile in the focal plane. For this design, the high-contrast plateau is $2\times10^{-7}$ the PSF peak, while the discovery space ranges from $4$ to $15$ $\lambda$/D.}
\end{figure}

The Shaped Pupil Coronagraph (SP hereafter) acts on the shape of the telescope pupil in order to enhance contrast in the focal plane\cite{Alexis}. The re-shaping process is obtained introducing a binary mask in the pupil plane. These masks are generated with an algorithm that iteratively converges to the pupil shape with the maximum throughput that satisfies the desired requirements in terms of the high-contrast region morphology and depth. In the process, any telescope aperture geometry can be considered. Figure~\ref{fig:SP_example} shows an example of binary mask, together with the generated high-contrast region in the focal plane. All the light falling out of this region is blocked introducing a hard edge mask in combination with a field stop. Finally, a Lyot stop is placed in the second pupil plane to block residual light diffracted at the edge of the pupil (as in the classical Lyot configuration).\\
We foresee to implement three SP in SHARK-NIR, one designed for moderate contrast but very small IWA and the other two optimized to be used in combination to achieve very high contrast on a quite small IWA.\\ 

\subsubsection{Four Quadrant Phase Mask}
\label{sec:FQPM}
This coronagraph suppresses on-axis starlight by means of a phase mask in the focal plane. The mask ideally divides the FP into four quadrants to apply in two of them on one diagonal a $\pi$ phase shift. Provided that the image of the star is perfectly centered on the common vertex of the quadrants, then the four outcoming beams combine destructively at infinity and the stellar light in the downstream pupil plane is totally rejected outside of the pupil area\cite{FQPM}. This light is then easily blocked by means of a Lyot stop.\\The FQPM represents an intriguing solution for SHARK-NIR, ensuring a way to further increase the detection potentialities of the instrument at small angular separations in regimes of very high SR and moderately small residual vibrations.

\subsubsection{Vortex coronagraph}
\label{sec:VC}
Like the FQPM, the vortex acts on the phase of the wavefront in the focal plane. Here the phase shift, while being constant with the distance from the center of the field, is a continuous function of the azimuthal coordinate, forming the so called 'phase ramp'. It has been shown in the laborary and on-sky that with this phase modification the light in the downstream pupil plane is completely removed from the geomerical area of pupil itself and thus can be easily blocked with a Lyot stop. This optical system has been proven, both on laboratory and on-sky, to achieve very high stellar suppression\cite{VORTEX_KECK}. Simulations with our test bench show that this technique yields very similar performance to the FQPM, but it is slightly more sensitive to low-order aberrations and misalignments.

\subsubsection{Apodized Pupil Lyot Coronagraph}
\label{sec:APLC}
\noindent APLC is the coronagraph adopted in top-level high contrast imagers such as SPHERE and GPI\cite{APLC}. While the SP creates high contrast with a single component (the apodizer), the basic idea behind APLC is to exploit the diffractive properties of the combination of apodizer, FP mask and Lyot stop. This technique thus relies on the joint optimization of the three. For a series of reasons related to instrument opto-mechanical constraints, simulations showed that this technique can not guarantee the same performance as other solutions and thus it is not foressen for the instrument.\\

\section{Applications}
\label{sec:results}
In this section we show a few situations which exemplify how the test bench can be used and which kind of analysis it allows to perform.\\Figure 2 shows, for each of the techniques implemented, the PSF out of the coronagraph, the PSF with the coronagraph and the raw contrast radial profile. Raw contrast here is simply the intensity in the coronagraphic PSF normalized to the peak of the out-of-coronagraph PSF. The conditions of the simulation are summarized in table 1. Intensities are shown in the same logarithmic scale (the scale is different between first and second column because of the different dynamic range) and the spatial extent of the image plane is also identical. Each coronagraph has its peculiar features: the PSF of the SP, for example, is determined by the presence of both an occulter and a field stop in the intermediate focal plane, while in the FQPM image the lines of phase discontinuity are clearly visible. The AO control radius appears clearly as a ring of bright speckles causing the bump at 550 mas in the radial profiles. The small circular ring close to the star in both FQPM and vortex PSF is due to telescope vibrations.\\ 

\noindent Figure 3 shows four images obtained integrating for one second with the Gaussian Lyot coronagraph in different seeing conditions, from 0.4'' to 1.0''. The star is bright (R=8), while vibrations are small (3 mas rms). It is clearly visible how a bad seeing enhances speckle brightness inside the AO controlled region. These speckles are distributed in the vertical direction, which is the one of the wind in simulations. Figure 4 reports the corresponding raw contrast profiles. The difference in contrast can be quantified as roughly half a magnitude per seeing step.\\

\noindent Another example of a possible application of the test bench is shown in figure 5. Here we compare three images obtained with a Four-Quadrant Phase-Mask coronagraph introducing a growing amount of vibrations. The difference in light-rejection capability of the coronagraph is clearly visible as vibration power increases: tip-tilt aberrations cause a characteristic light ring to form around the star. As vibration power increases, secondary rings start to appear at increasing angular separations. 

\section{CONCLUSIONS}
\label{sec:CON}
In this paper we have presented a Fresnel simulator developed in the framework of the SHARK-NIR project, the second generation high-contrast imager of the Large Binocular Telescope. The simulator is conceived as a test bench: it implements several coronagraphic techniques and can generate images in a wide range of operative conditions. Several sources of optical aberrations are introduced, with particular care on realistically reproducing the LBT environment. The examples of possible applications shown here are part of a more comprehensive tradeoff study that led to the definition of an optimal suite of coronagraphic designs for the instrument.

\newpage

\bibliography{bib} 

\begin{thebibliography}{10}

\bibitem{PROPER}
J.~E. Krist, ``{PROPER}: an optical propagation library for {IDL},'' {\em
  Society of Photo-Optical Instrumentation Engineers (SPIE) Conference
  Series}~{\bf 6675}, Sept. 2007.

\bibitem{FLAO}
S.~Esposito, A.~Riccardi, and L.~Fini, ``{LBT AO} on-sky results,'' {\em Second
  International Conference on Adaptive Optics for Extremely Large Telescopes} ,
  2011.

\bibitem{M2}
A.~Riccardi, M.~Xompero, R.~Briguglio, F.~Quiros-Pacheco, L.~Busoni, L.~Fini,
  A.~Puglisi, S.~Esposito, C.~Arcidiacono, E.~Pinna, P.~Ranfagni, P.~Salinari,
  G.~Brusa, R.~Demers, R.~Biasi, and D.~Gallieni, ``{T}he adaptive secondary
  mirror for the {L}arge {B}inocular {T}elescope: optical acceptance test and
  preliminary on-sky commissioning results,'' {\em Society of Photo-Optical
  Instrumentation Engineers (SPIE) Conference Series}~{\bf 7736}, Jul. 2010.

\bibitem{PIRAMIDE}
R.~Ragazzoni and J.~Farinato, ``{S}ensitivity of a pyramidic {W}ave {F}ront
  sensor in closed loop {A}daptive {O}ptics,'' {\em Astronomy and
  Astrophysics}~{\bf 350}, pp.~L23--L26, Oct. 1999.

\bibitem{PASSATA}
G.~Agapito, A.~Puglisi, and S.~Esposito, ``{PASSATA}: object oriented numerical
  simulation software for adaptive optics,'' {\em Society of Photo-Optical
  Instrumentation Engineers (SPIE) Conference Series}~{\bf 9909}, p.~9, Jul.
  2016.

\bibitem{FORERUNNER}
F.~Pedichini, M.~Stangalini, A.~Ambrosino, A.~Puglisi, E.~Pinna, V.~Bailey,
  L.~Carbonaro, M.~Centrone, J.~Christou, S.~Esposito, J.~Farinato, F.~Fiore,
  E.~Giallongo, J.~M. Hill, P.~M. Hinz, and L.~Sabatini, ``{H}igh {C}ontrast
  {I}maging in the {V}isible: {F}irst {E}xperimental {R}esults at the {L}arge
  {B}inocular {T}elescope,'' {\em The Astronomical Journal}~{\bf 154}(2), p.~5,
  Aug. 2017.

\bibitem{Alexis}
A.~Carlotti, R.~Vanderbei, and N.~J. Kasdin, ``Optimal pupil apodizations of
  arbitrary apertures for high-contrast imaging,'' {\em Optics Express}~{\bf
  19}, p.~26796, 2011.

\bibitem{FQPM}
D.~Rouan, P.~Riaud, A.~Boccaletti, Y.~Clénet, and A.~Labeyrie, ``The
  {F}our-{Q}uadrant {P}hase-{M}ask {C}oronagraph. i. {P}rinciple,'' {\em
  Society of Photo-Optical Instrumentation Engineers (SPIE) Conference
  Series}~{\bf 112}, p.~8, Nov. 2000.

\bibitem{VORTEX_KECK}
E.~Serabyn, E.~Huby, K.~Matthews, D.~Mawet, O.~Absil, B.~Femenia,
  P.~Wizinowich, M.~Karlsson, M.~Bottom, R.~Campbell, B.~Carlomagno,
  D.~Defrère, C.~Delacroix, P.~Forsberg, C.~Gomez~Gonzalez, S.~Habraken,
  A.~Jolivet, K.~Liewer, S.~Lilley, P.~Piron, M.~Reggiani, J.~Surdej, H.~Tran,
  E.~Vargas~Catalán, and O.~Wertz, ``The {W}. {M}. {K}eck {O}bservatory
  {I}nfrared {V}ortex {C}oronagraph and a {F}irst {I}mage of {HIP} 79124 {B},''
  {\em The Astronomical Journal}~{\bf 153}, p.~7, Jan. 2017.

\bibitem{APLC}
C.~Aime, R.~Soummer, and A.~Ferrari, ``{T}otal coronagraphic extinction of
  rectangular apertures using linear prolate apodizations,'' {\em Astronomy and
  Astrophysics}~{\bf 389}, pp.~334--344, Jul. 2002.

\end{thebibliography}
\bibliographystyle{spiebib} 

\newpage

\begin{table}[h]\centering \renewcommand\arraystretch{1.2} 
\label{tab:parameters}
\begin{tabular}{l c}
\rowcolor{gray!10} Wavelength  & $1.6$ $\mu$m\\
R mag  & 8 \\
\rowcolor{gray!10} H mag  & 6 \\
Seeing  & 0.4'' \\
\rowcolor{gray!10} DIT & 1 s \\
Residual jitter & 10 mas rms\\
\rowcolor{gray!10} NCPA & 30 nm\\
\end{tabular}
\vspace{0.3cm}
\caption{Parameters used for simulating the images shown in figure 2.}
\end{table}

\vspace{2.5cm} 

\begin{figure}[h]
\label{fig:coro1}
\centering
\begin{multicols}{3}
{\includegraphics[width=\linewidth]{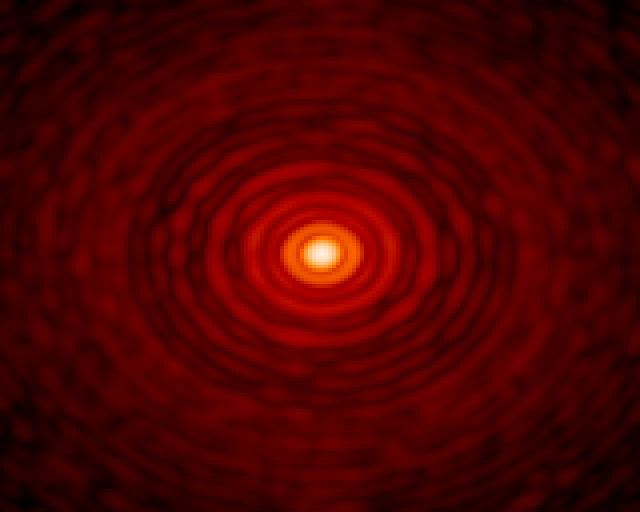}}\par
\includegraphics[width=\linewidth]{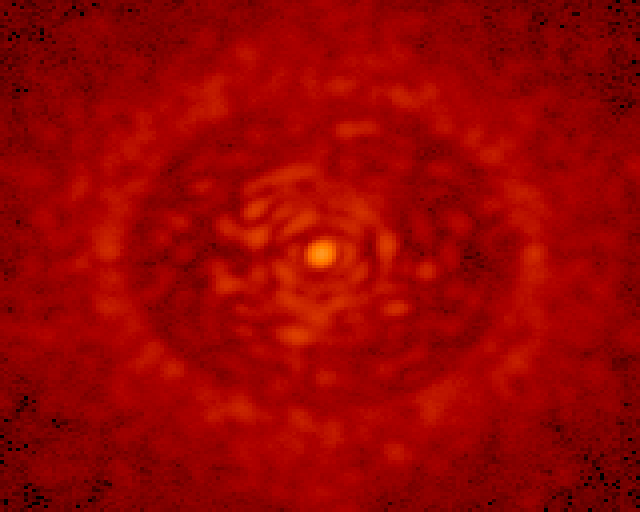}\par
{\includegraphics[width=\linewidth,height=0.82\linewidth]{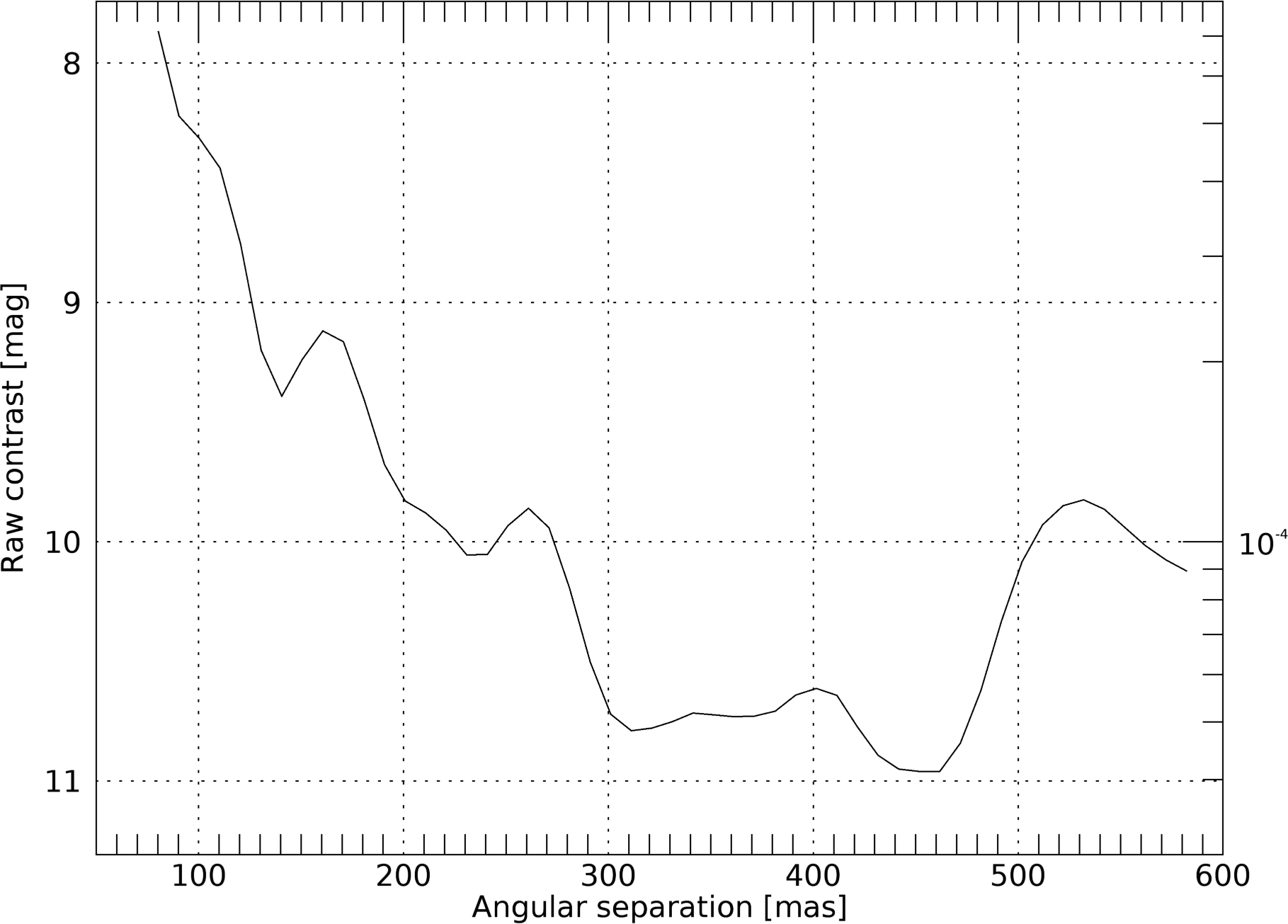}}\par 
\end{multicols}

\begin{multicols}{3}
{\includegraphics[width=\linewidth]{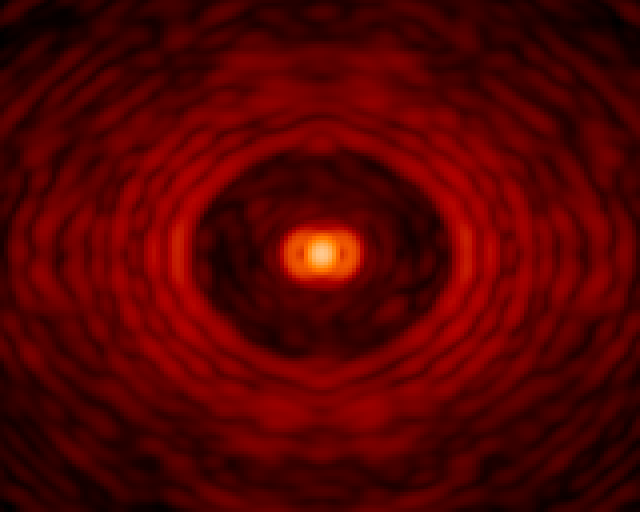}}\par
\includegraphics[width=\linewidth]{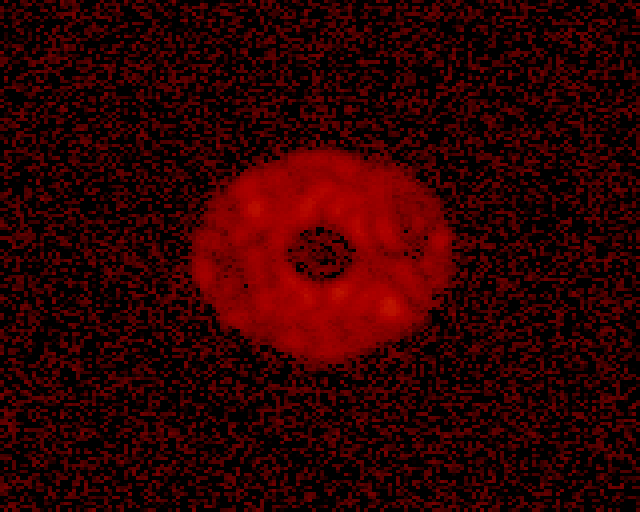}\par
{\includegraphics[width=\linewidth,height=0.82\linewidth]{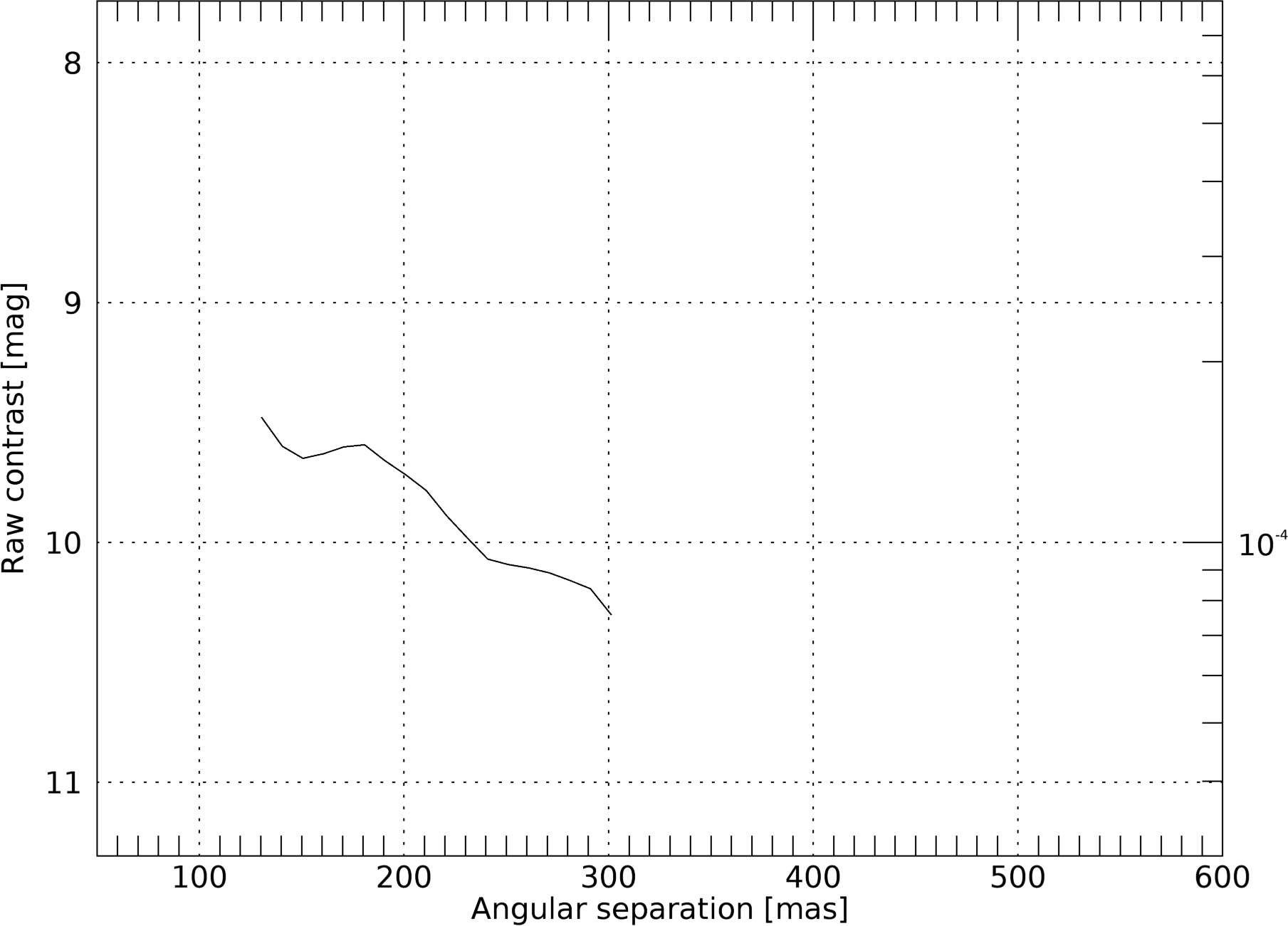}}\par 
\end{multicols}

\begin{multicols}{3}
{\includegraphics[width=\linewidth]{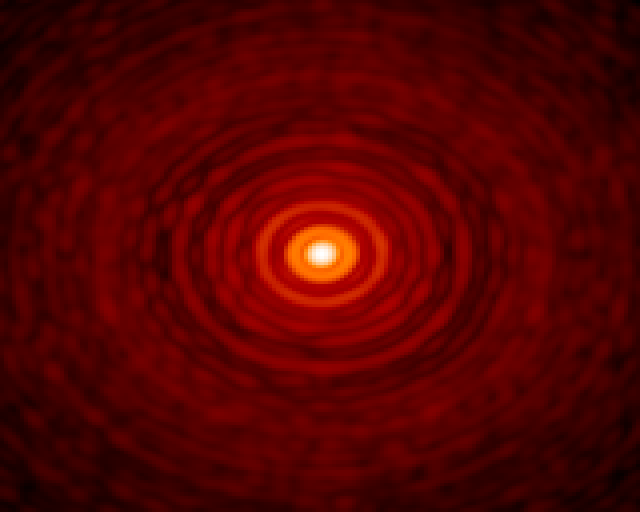}}\par
\includegraphics[width=\linewidth]{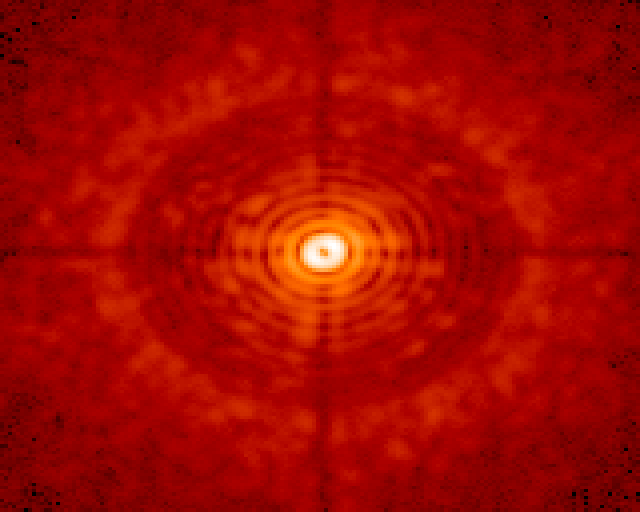}\par
{\includegraphics[width=\linewidth,height=0.82\linewidth]{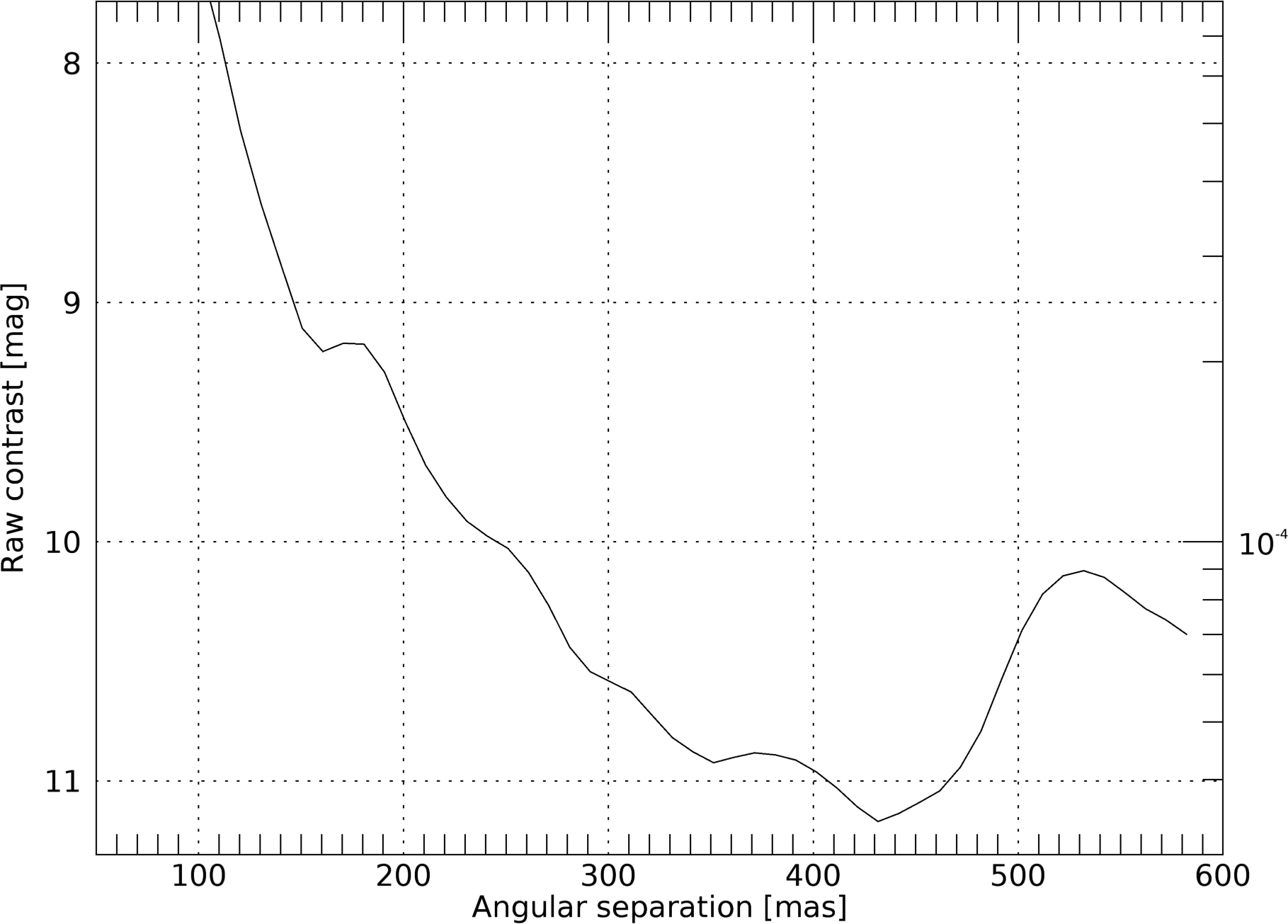}}\par 
\end{multicols}

\begin{multicols}{3}
{\includegraphics[width=\linewidth]{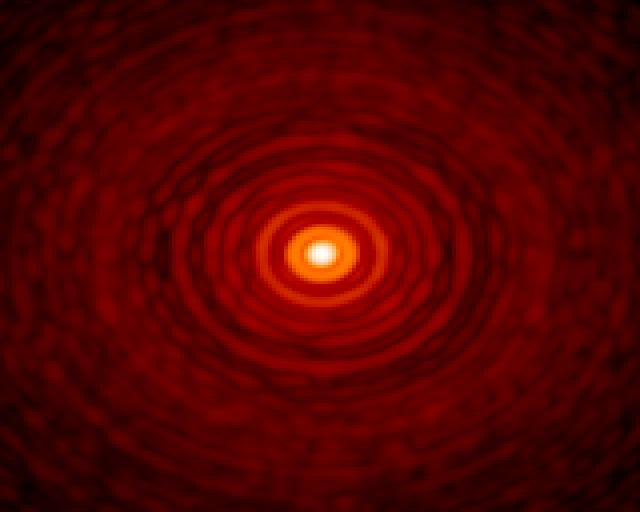}}\par
\includegraphics[width=\linewidth]{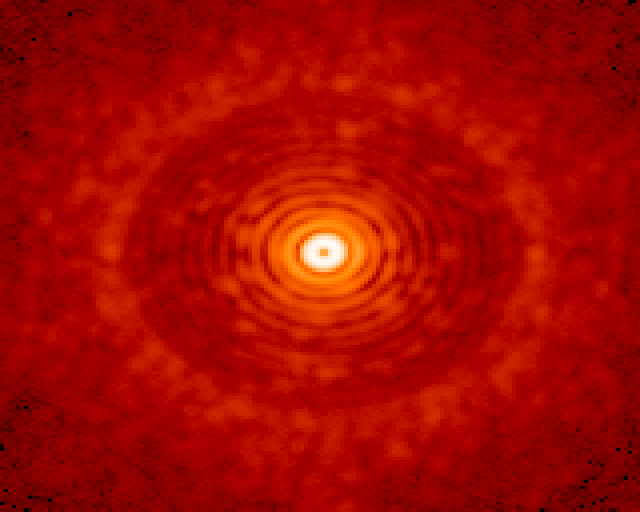}\par
{\includegraphics[width=\linewidth,height=0.82\linewidth]{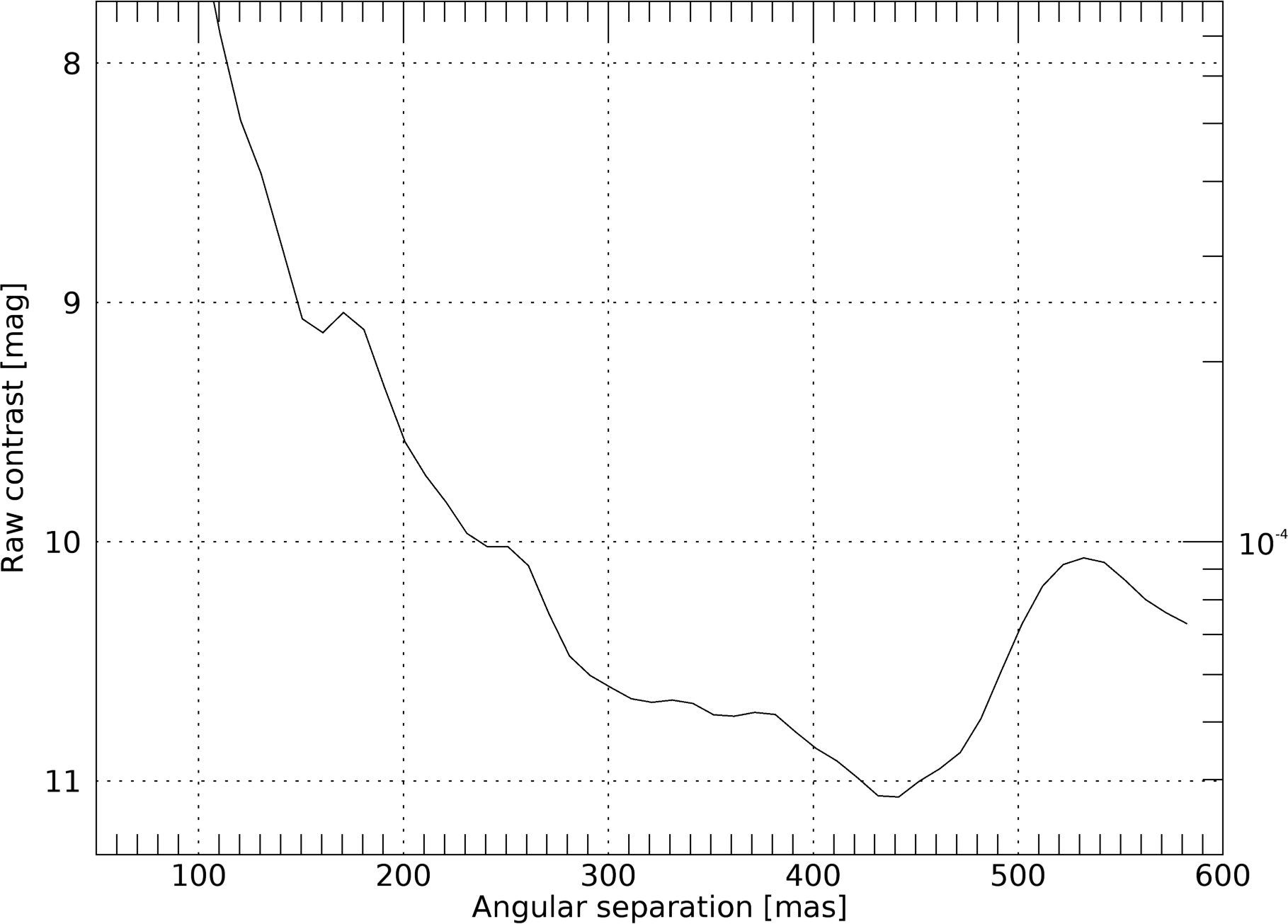}}\par 
\end{multicols}

\end{figure}	

\begin{figure}[h] \label{fig:coro2}
\centering
\begin{multicols}{3}
{\includegraphics[width=\linewidth]{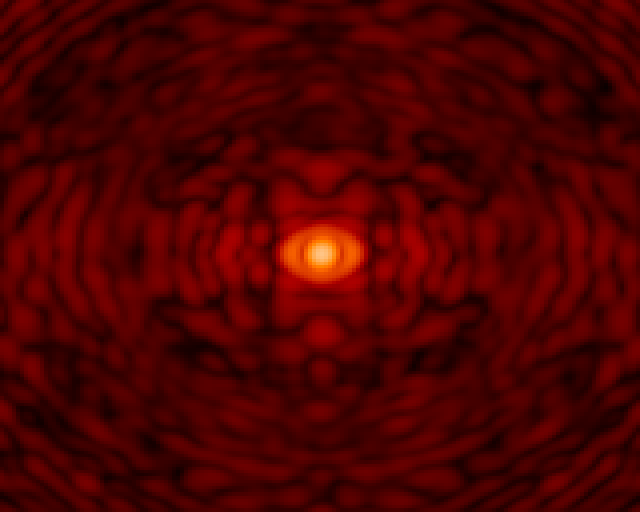}}\par
\includegraphics[width=\linewidth]{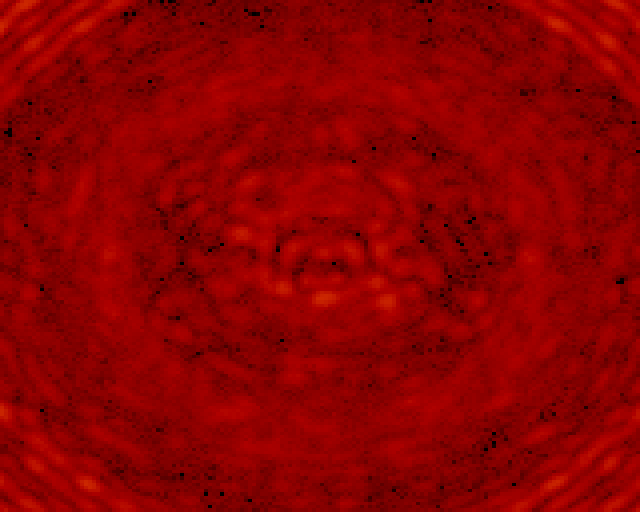}\par
{\includegraphics[width=\linewidth,height=0.82\linewidth]{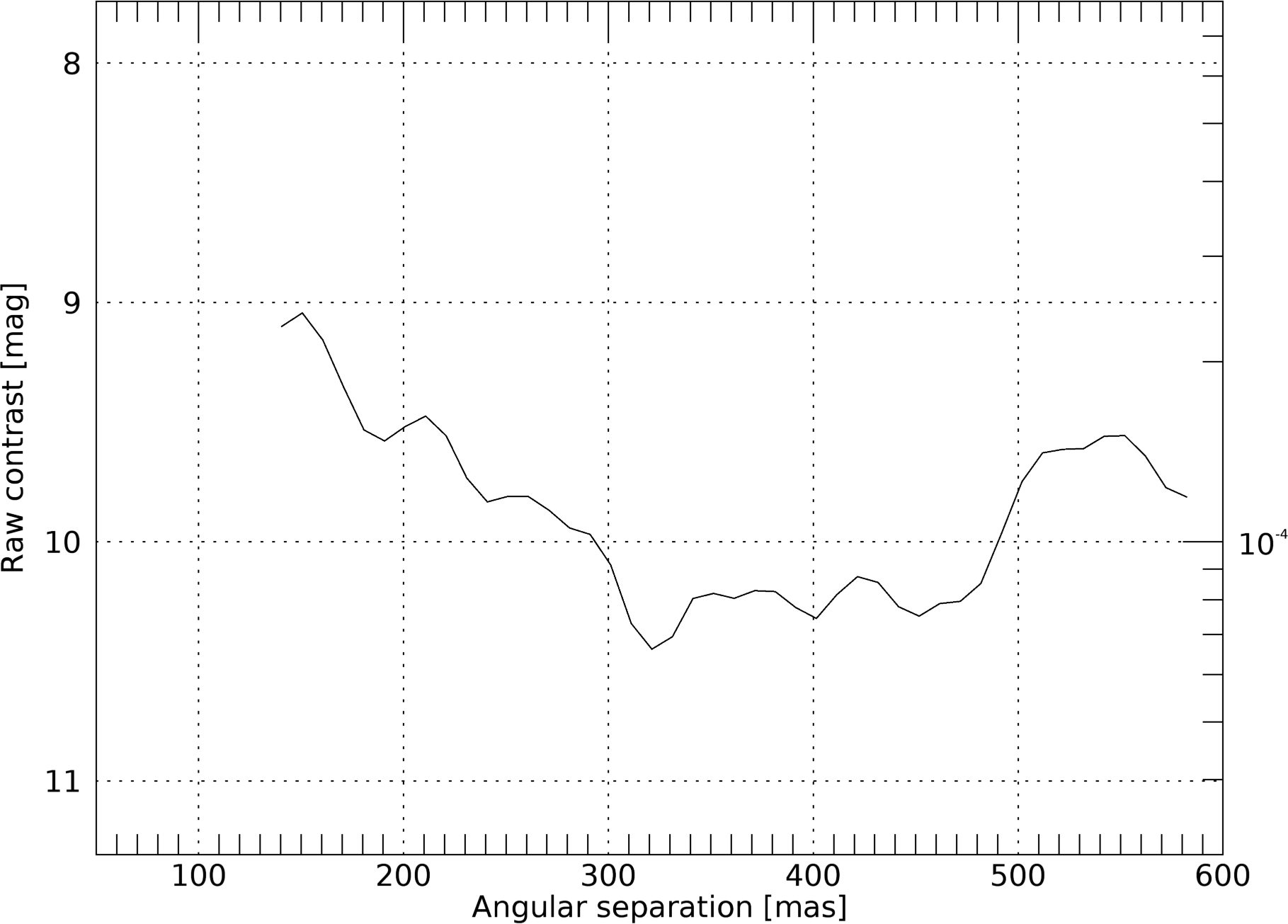}}\par 
\end{multicols}
\caption{PSF out of the coronagraph, PSF with the coronagraph and corresponding raw contrast radial profile for each of the techniques implemented in the virtual test bench. In order: Gaussian Lyot, Shaped Pupil, Four-Quadrant Phase-Mask, Vortex and Apodized Pupil Lyot Coronagraph. Intensities are shown in the same logarithmic scale.}
\end{figure}

\begin{figure}
\label{fig:seeing}
\centering
{\includegraphics[width=.42\columnwidth]{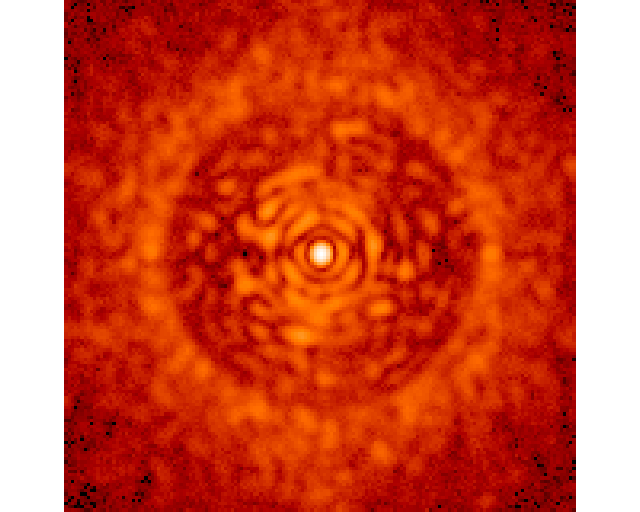}} \quad
\includegraphics[width=.42\columnwidth]{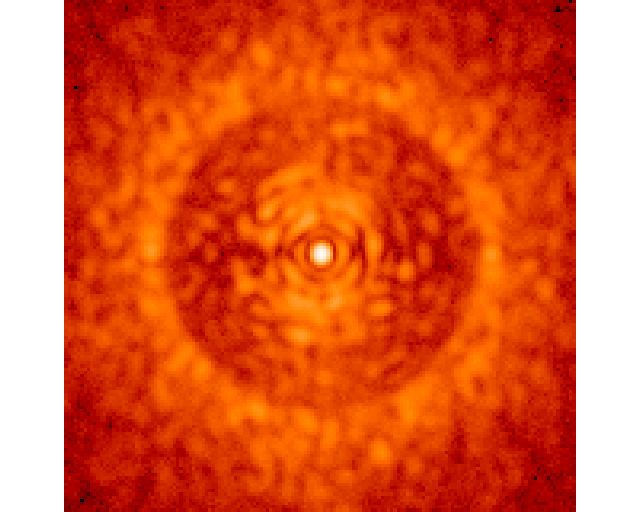} \\ \vspace{0.1cm}
{\includegraphics[width=.42\columnwidth]{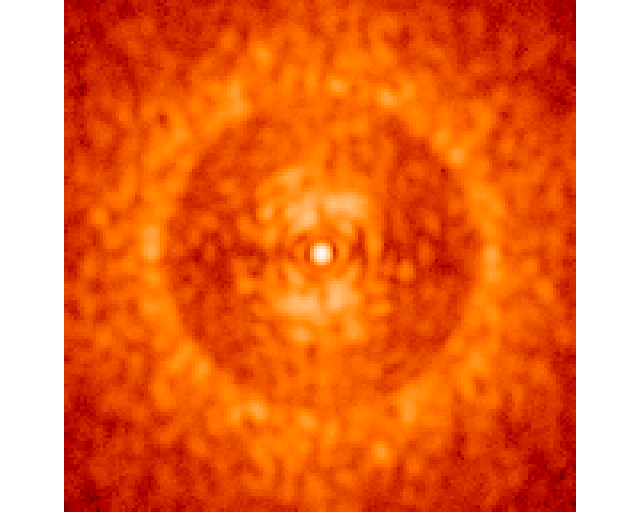}} \quad
\includegraphics[width=.42\columnwidth]{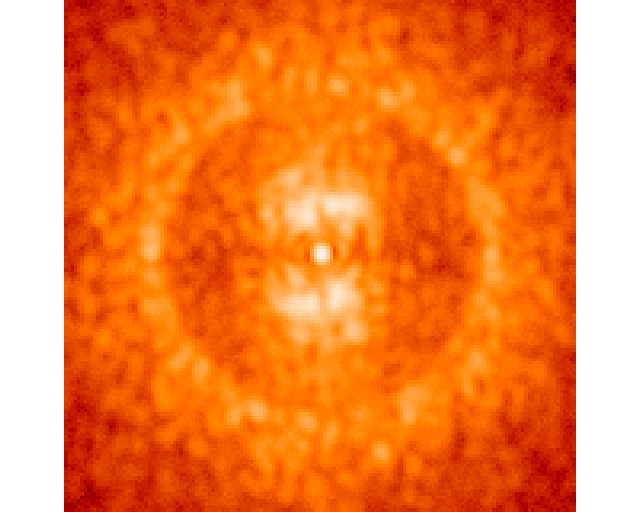} \\  \vspace{0.3cm}
\caption{Images obtained integrating after 1 second of integration and a Gaussian Lyot coronagraph in different seeing conditions: 0.4'' (top left), 0.6'' (top right), 0.8'' (bottom left) and 1.0'' (bottom right). Intensities are displayed in the same logarithmic scale. As the seeing get worse, bright speckles appear inside the AO control radius in the direction of the wind beacuse of the AO lag error.}
\end{figure}

\begin{figure}\label{fig:seeing_r}
\centering
\includegraphics[scale=0.90]{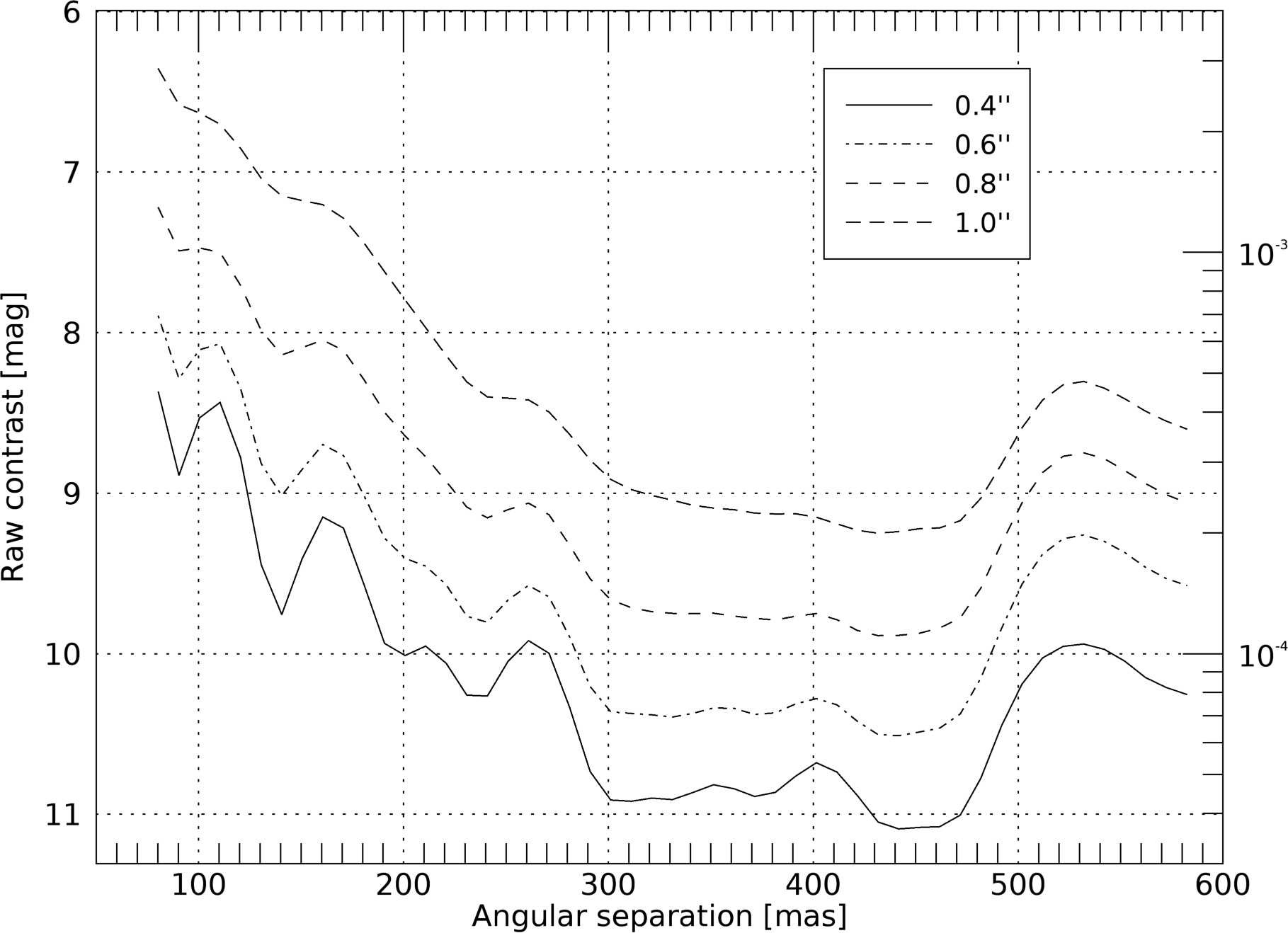}  \vspace{0.3cm}
\caption{Raw contrast profiles obtained using a Gaussian Lyot coronagraph in different seeing conditions. There is a costant loss of about half a magnitude at all angular separations between the curves as seeing get worse.}
\end{figure}

\vspace{1 cm}

\begin{figure}
\label{fig:jitter}
\centering
{\includegraphics[width=.45\columnwidth]{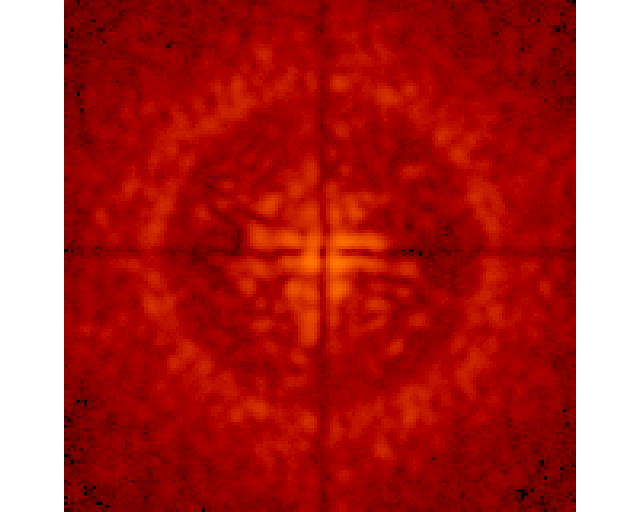}} \quad
\includegraphics[width=.45\columnwidth]{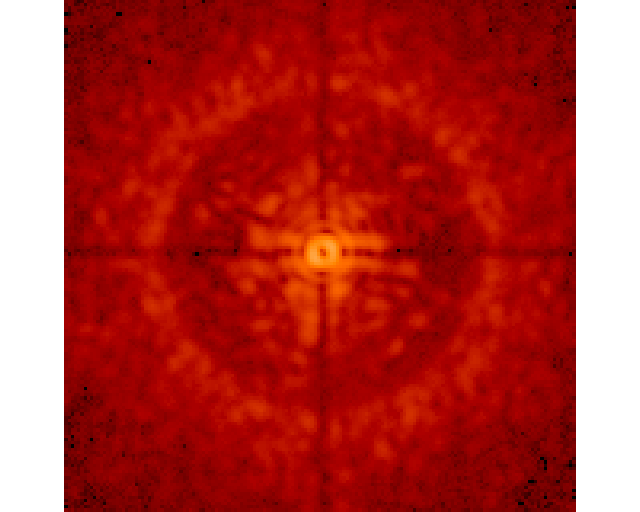} \\ \vspace{0.3cm}
{\includegraphics[width=.45\columnwidth]{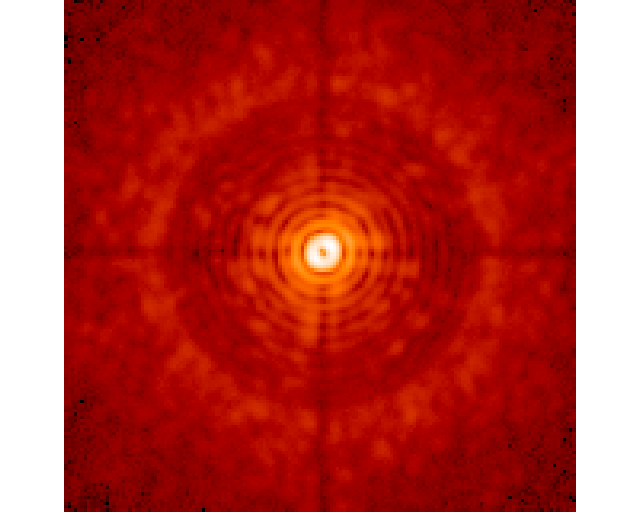}} \vspace{0.3cm}
\caption{Images obtained after 1 second of integration with a Four-Quadrant Phase-Mask coronagraph for different values of telescope vibrations: no vibrations (top left), 3 mas rms (top right) and 10 mas rms (bottom). Intensities are displayed in the same logarithmic scale. The difference in light-rejection capability of the coronagraph is clearly visible. Tip-Tilt aberration causes a bright ring to form very close to the star. As vibration power increases, secondary rings further away from the star become to appear.}
\end{figure}

\end{document}